\documentclass{article}

\usepackage{times}
\usepackage{graphicx} 
\usepackage{subfigure} 

\usepackage{natbib}

\usepackage{algorithm}
\usepackage{algorithmic}

\usepackage{hyperref}

\usepackage{amsmath}

\newcommand{\pd}[2]{\frac{\partial #1}{\partial #2}}

\newcommand{\mb}{\mathbf}
\newcommand{\mc}{\mathcal}

\usepackage[accepted]{icml2014}

\icmltitlerunning{Hamiltonian Monte Carlo Without Detailed Balance}

\begin{document} 

\twocolumn[
\icmltitle{Hamiltonian Monte Carlo Without Detailed Balance}

\icmlauthor{Jascha Sohl-Dickstein}{jascha@stanford.edu}
\icmladdress{Stanford University, Palo Alto.  Khan Academy, Mountain View}
\icmlauthor{Mayur Mudigonda}{mudigonda@berkeley.edu}
\icmladdress{Redwood Institute for Theoretical Neuroscience,
            University of California at Berkeley}
\icmlauthor{Michael R. DeWeese}{deweese@berkeley.edu}
\icmladdress{Redwood Institute for Theoretical Neuroscience,
            University of California at Berkeley}

\icmlkeywords{boring formatting information, machine learning, ICML}

\vskip 0.3in
]

\begin{abstract} 
We present a method for performing Hamiltonian Monte Carlo that largely eliminates sample rejection for typical hyperparameters.  In situations that would normally lead to rejection, instead a longer trajectory is computed until a new state is reached that can be accepted.  This is achieved using Markov chain transitions that satisfy the fixed point equation, but do not satisfy detailed balance.  The resulting algorithm significantly suppresses the random walk behavior and wasted function evaluations that are typically the consequence of update rejection.  We demonstrate a greater than factor of two improvement in mixing time on three test problems.  We release the source code as Python and MATLAB packages.
\end{abstract}

\section{Introduction}

High dimensional and otherwise computationally expensive probabilistic models are of increasing importance for such diverse tasks as modeling the folding of proteins \cite{Schutte1999}, the structure of natural images \cite{Culpepper2011}, or the activity of networks of neurons \cite{Cadieu2010}.

Sampling from the described distribution is typically the bottleneck when working with these probabilistic models.  Sampling is commonly required when training a probabilistic model, when evaluating the model's performance, when performing inference, and when taking expectations \cite{MacKay2003a}.  Therefore, work {that} improves sampling is fundamentally important.

The most common way to guarantee that a sampling algorithm converges to the correct distribution is via a concept known as detailed balance.  Sampling algorithms based on detailed balance are powerful because they allow samples from any target distribution to be generated from almost any proposal distribution, using for instance Metropolis-Hastings acceptance criteria \cite{Hastings:1970p13277}. 
However, detailed balance also suffers from a critical flaw.  By definition forward and reverse transitions occur with equal probability under detailed balance, and samplers that obey detailed balance go backwards exactly as often as they go forwards.  The state space is thus explored via a random walk over distances longer than those traversed by a single draw from the proposal distribution, and the number of steps to traverse the state is quadratic in the size of the state space. Samplers that violate detailed balance can mix more rapidly
\cite{diaconis2000analysis,chen1999lifting,sun2010improving,suwa2010markov,turitsyn2011irreversible,Ichiki2013,hukushima2013irreversible,ohzeki2013acceleration,kondo2013enhanced,bierkens2014non}.

The current state-of-the-art sampling algorithm for probability distributions with continuous state spaces is Hamiltonian Monte Carlo (HMC) \cite{Duane1987,Neal:HMC}.  By extending the state space to include auxiliary momentum variables, and then using Hamiltonian dynamics to traverse long iso-probability contours in this extended state space, HMC is able to move long distances in state space in a single update step.  However, HMC still relies on detailed balance to accept or reject steps, and as a result still behaves like a random walk -- just a random walk with a longer step length.  Previous attempts to address this have combined multiple Markov steps that individually satisfy detailed balance into a composite step that does not \cite{Horowitz1991a}, with limited success \cite{Kennedy1991}.  

The No-U-Turn Sampler (NUTS) sampling package \cite{Hoffman2011a} and the windowed acceptance method of \cite{Neal1994b} both consider Markov transitions within a set of discrete states generated by repeatedly simulating Hamiltonian dynamics.  NUTS generates a set of candidate states around the starting state by running Hamiltonian dynamics forwards and backwards until the trajectory doubles back on itself, or a slice variable constraint is violated.  It then chooses a new state at uniform from the candidate states.  In windowed acceptance, a transition is proposed between a window of states at the beginning and end of a trajectory, rather than the first state and last state.  Within the selected window, a single state is then chosen using Boltzmann weightings.  Both NUTS and the windowed acceptance method rely on detailed balance to choose the candidate state from the discrete set.

Here we present a novel discrete representation of the HMC state space and transitions.  Using this representation, we derive a method for performing HMC without relying on detailed balance, by directly satisfying the fixed point equation restricted to the discrete state space.  As a result, random walk behavior in the sampling algorithm is greatly reduced, and the mixing rate of the sampler is substantially improved.

\section{Sampling}

We begin by briefly reviewing some key concepts related to sampling.  The goal of a sampling algorithm is to draw characteristic samples $\mb x \in \mc R^N$ from a target probability distribution $p\left( \mb x \right)$.  Without loss of generality, we will assume that $p\left( \mb x \right)$ is determined by an energy function $E\left(\mb x\right)$,
\begin{align}
p\left(\mb x \right)
	&=
		\frac{1}{Z}\exp\left(-E\left(\mb x\right)\right)
.
\end{align}

\subsection{Markov Chain Monte Carlo}

Markov Chain Monte Carlo (MCMC) \cite{Neal1993} is commonly used to sample from probabilistic models.  In MCMC a chain of samples is generated by repeatedly drawing new samples $\mb x'$ from a conditional probability distribution $T\left(\mb x' | \mb x \right)$, where $\mb x$ is the previous sample.  Since $T\left(\mb x' | \mb x \right)$ is a probability 
{density}
over $\mb x'$, $\int  T\left( \mb x' | \mb x \right) d\mb x' = 1$ and $T\left( \mb x' | \mb x \right) \geq 0$.

\subsection{Fixed Point Equation}

An MCMC algorithm must satisfy two conditions in order to generate samples from the target distribution $p\left( \mb x \right)$.  The first is mixing, which {requires} that repeated application of $T\left(\mb x' | \mb x \right)$ must eventually explore the full state space of $p\left( \mb x \right)$.  The second condition is that the target distribution $p\left( \mb x \right)$ must be a fixed point of $T\left(\mb x' | \mb x \right)$.  This second condition can be expressed by the fixed point equation,
\begin{align}
\int p\left( \mb x \right) T\left( \mb x' | \mb x \right) d\mb x &= p\left( \mb x' \right)
,
\label{eq fixed}
\end{align}
which {requires} that when $T\left( \mb x' | \mb x \right)$ acts on $p\left( \mb x \right)$, the resulting distribution is unchanged.

\subsection{Detailed Balance}

Detailed balance is the most common way of guaranteeing that the Markov transition distribution $T\left( \mb x' | \mb x \right)$ satisfies the fixed point equation {(Equation~\ref{eq fixed})}.  Detailed balance guarantees that if samples are drawn from the equilibrium distribution $p\left(\mb x\right)$, then for every pair of states $\mb x$ and $\mb x'$ the probability of transitioning from state $\mb x$ to state $\mb x'$ is identical to that of transitioning from state $\mb x'$ to $\mb x$,
\begin{align}
p\left( \mb x \right) T\left( \mb x' | \mb x \right) 
	&=
p\left( \mb x' \right) T\left( \mb x | \mb x' \right) 
.
\label{eq detailed}
\end{align}
By substitution for $T\left( \mb x' | \mb x \right)$ in the left side of Equation \ref{eq fixed}, it can be seen that if Equation \ref{eq detailed} is satisfied, then the fixed point equation is also satisfied.

An appealing aspect of detailed balance is that a transition distribution satisfying it can be easily constructed from {nearly} any proposal distribution, using Metropolis-Hastings acceptance/rejection rules \cite{Hastings:1970p13277}.  A primary drawback of detailed balance, and of Metropolis-Hastings, is that the resulting Markov chains always engage in random walk behavior, since by definition detailed balance depends on forward and reverse transitions happening with equal probability.
 
 The primary advance in this paper is demonstrating how HMC sampling can be performed without resorting to detailed balance.

\section{Hamiltonian Monte Carlo}

Hamiltonian Monte Carlo (HMC) 
can traverse long distances in state space with single Markov transitions.  It does this by extending the state space to include auxiliary momentum variables, and then simulating Hamiltonian dynamics to move long distances along iso-probability contours in the expanded state space.

\subsection{Extended state space}

The state space is extended by the addition of momentum variables $\mb v \in \mc R^N$, with identity-covariance Gaussian distribution,
\begin{align}
p\left( \mb v \right) 
	&=
\left(2\pi\right)^{-\frac{N}{2}}
\exp\left( -\frac{1}{2}\mb v^T \mb v \right)
.
\label{eq v}
\end{align}
We refer to the combined state space of $\mb x$ and $\mb v$ as $\mb \zeta$, such that $\zeta = \left\{ \mb x, \mb v \right\}$.  The corresponding joint distribution is
\begin{align}
p\left( \mb \zeta \right) 
	&=
p\left( \mb x, \mb v \right) 
	=
p\left( \mb x \right) p\left( \mb v \right) 
	=
\frac{\left(2\pi\right)^{-\frac{N}{2}}}{Z}
\exp\left( -H\left( \zeta \right) \right)
,
\label{eq joint} \\
H\left( \mb \zeta \right) &= H\left( \mb x, \mb v \right) = E\left( \mb x \right) + \frac{1}{2}\mb v^T \mb v
.
\end{align}
$H\left( \mb \zeta \right)$ has the same form as total energy in a physical system, where $E\left( \mb x \right)$ is the potential energy for position $\mb x$ and $\frac{1}{2}\mb v^T \mb v$ is the kinetic energy for momentum $\mb v$ (mass is set to one). 

In HMC samples from $p\left( \mb x \right)$ are generated by drawing samples from the joint distribution $p\left( \mb x, \mb v \right)$, and retaining only the $\mb x$ variables as samples from the desired distribution.

\subsection{Hamiltonian dynamics}

Hamiltonian dynamics govern how physical systems evolve with time.  It might be useful to imagine the trajectory of a skateboarder rolling in an empty swimming pool.  As she rolls downwards she exchanges potential energy for kinetic energy, and the magnitude of her velocity increases.  As she rolls up again she exchanges kinetic energy back for potential energy.  In this fashion she is able to traverse long distances across the swimming pool, while at the same time maintaining constant total energy over her entire trajectory.

In HMC, we treat $H\left( \mb \zeta \right)$ as the total energy of a physical system, with spatial coordinate $\mb x$, velocity $\mb v$, potential energy $E\left( \mb x \right)$, and kinetic energy $\frac{1}{2}\mb v^T \mb v$.  In an identical fashion to the case of the skateboarder in the swimming pool, running Hamiltonian dynamics on this system traverses long distances in $\mb x$ while maintaining constant total energy $H\left( \mb \zeta \right)$.  By Equation \ref{eq joint}, moving along a trajectory with constant energy is identical to moving along a trajectory with constant probability density.

Hamiltonian dynamics can be run exactly in reverse by reversing the velocity vector.  They also preserve volume in $\mb \zeta$.  As we will see, all these properties together mean that Hamiltonian dyamics can be used to propose update steps {that} move long distances in state space while retaining high acceptance probability.

%

\subsection{Operators}
\label{sec op}
\begin{figure}
\begin{center}
\hspace{-10mm}
\parbox[c]{0.95\linewidth}{
\begin{tabular}[b]{cc}
\begin{tabular}[b]{l}
	\includegraphics[width=0.45\linewidth]{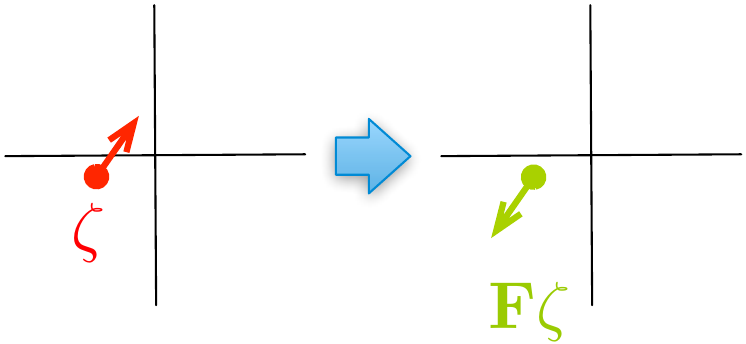}\\
	\includegraphics[width=0.475\linewidth]{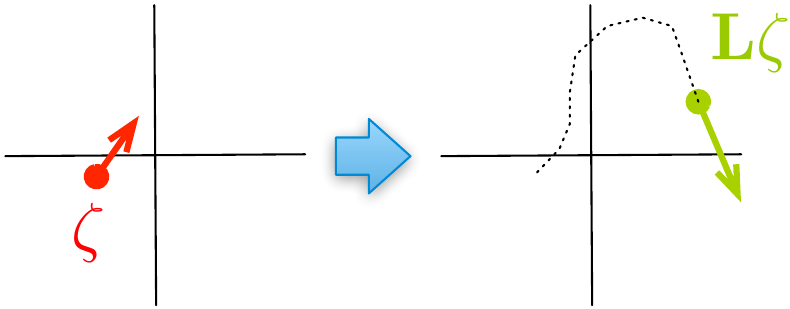}\\
	\includegraphics[width=0.45\linewidth]{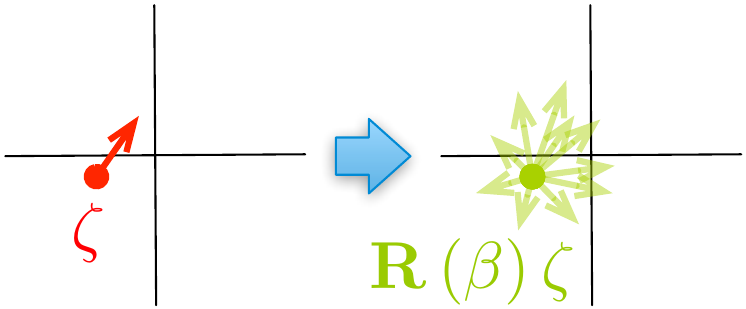}\\(a)
\end{tabular}
&
\begin{tabular}[b]{l}
	\includegraphics[width=0.45\linewidth]{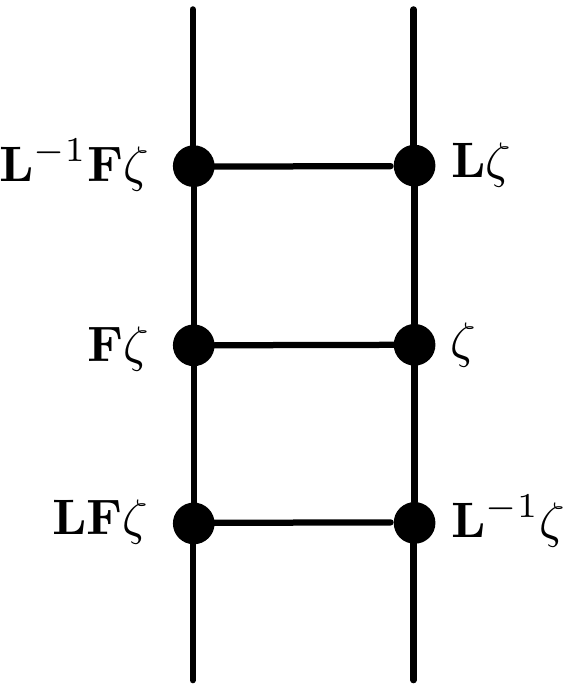}\\(b)
\end{tabular}
\end{tabular}
} 
\end{center}
\caption{
{\em (a)} The action of operators involved in Hamiltonian Monte Carlo (HMC).  The base of each {red or green arrow} represents the position $\mb x$, and the length and direction of {each of these arrows} represents the momentum $\mb v$.  The flip operator $\mb F$ reverses the momentum.  The leapfrog operator $\mb L$ approximately integrates Hamiltonian dynamics.  The trajectory taken by $\mb L$ is indicated by the dotted line.  
The randomization operator $\mb R\left(\beta\right)$ corrupts the momentum with an amount of noise {that} depends on $\beta$.  {\em (b)} The ladder of discrete states {that} are accessible by applying $\mb F$ and $\mb L$ starting at state $\mb \zeta$.  {Horizontal movement} on the ladder occurs by flipping the momentum, {whereas vertical movement} occurs by integrating Hamiltonian dynamics. 
}
\label{fig op}
\end{figure}

The Markov transitions from which HMC is constructed can be understood in terms of several operators acting on $\mb \zeta$.  These operators are illustrated in Figure \ref{fig op}a.  This representation of the actions performed in HMC, and the corresponding state space, is unique to this paper and diverges from the typical presentation of HMC.

\subsubsection{Momentum Flip}

The momentum flip operator $\mb F$ 
{reverses the direction of} the momentum.  It is its own inverse, leaves the total energy unchanged, and preserves volume in state space{:}
\begin{align}
\mb F \mb \zeta = \mb F \left\{\mb x, \mb v\right\} &= \left\{\mb x, -\mb v\right\}, \label{op F}
\\
\mb F^{-1}  \mb \zeta &= \mb F \mb \zeta,
\\
H\left( \mb F \mb \zeta \right) &=  H\left( \mb \zeta \right),
\\
\left| \det\left( \pd{
	\mb F \mb \zeta
	}{
		\zeta^T
	} \right)\right|
&= 1
.
\end{align}
The momentum flip operator $\mb F$ causes movement between the left and right sides of the state ladder in Figure \ref{fig op}b.

\subsubsection{Leapfrog Integrator}

Leapfrog, or St\"{o}rmer-Verlet, integration provides a discrete time approximation to Hamiltonian dynamics \cite{Hairer2003}.  The operator $\mb L\left( \epsilon, M \right)$ performs leapfrog integration for $M$ leapfrog steps with step length $\epsilon$.  For conciseness, $\mb L\left( \epsilon, M \right)$ will be written only as $\mb L$,
\begin{align}
\mb L \mb \zeta &= {\left\{ \parbox{15em}{The state resulting from $M$ steps of leapfrog integration of Hamiltonian dynamics with step length $\epsilon$.}\right.}
\end{align}
Like exact Hamiltonian dynamics, leapfrog dynamics are exactly reversible by reversing the velocity vector, and {they} also exactly preserve volume in state space.  $\mb L$ can be inverted by 
{reversing the sign of} the momentum, tracing out the reverse trajectory, and then 
{reversing the sign of} the momentum again so that it points in the original direction;
\begin{align}
& \hspace{.1in} \mb L^{-1} \mb \zeta =  \mb F \mb L \mb F  \mb \zeta
,
\\
&\left| \det\left( \pd{
	\mb L \mb \zeta
	}{
		\zeta^T
	} \right)\right|
= 1.
\end{align}
Unlike for exact dynamics, the total energy $H\left( \mb \zeta \right)$ is only approximately conserved by leapfrog integration, and the energy accumulates errors due to discretization.  This discretization error in the energy is the source of all rejections of proposed updates in HMC.

The leapfrog operator $\mb L$ causes movement up the right side of the state ladder in Figure \ref{fig op}b, and down the left side of the ladder.
%

\subsubsection{Momentum Randomization}

The momentum randomization operator $\mb R\left(\beta\right)$ mixes an amount of Gaussian noise determined by $\beta \in [0, 1]$ into the velocity vector,
\begin{align}
\mb R\left(\beta\right)\mb \zeta &=  \mb R\left(\beta\right)\left\{\mb x, \mb v\right\} = \left\{\mb x, \mb v' \right\}
, \\
\mb v' &= \mb v \sqrt{1 - \beta} + \mb n \sqrt{\beta}
, \\
\mb n & \sim N\left( \mb 0, \mb I\right)
.
\end{align}
Unlike the previous two operators, the momentum randomization operator is not deterministic.  $\mb R\left(\beta\right)$ is however a valid Markov transition operator for $p\left(\mb \zeta \right)$ on its own, in that it satisfies both Equation \ref{eq fixed} and Equation \ref{eq detailed}.

The momentum randomization operator $\mb R\left(\beta\right)$ causes movement off of the current state ladder and onto a new state ladder.

\subsection{Discrete State Space}
\label{sec discrete}
As illustrated in Figure \ref{fig op}b, the operators $\mb L$ and $\mb F$ generate a discrete state space ladder, with transitions only occurring between $\mb \zeta$ and three other states.  Note that every state on the ladder can be represented many different ways, depending on the series of operators used to reach it.  For instance, the state in the upper left of the figure pane can be written $\mb L^{-1} \mb F \mb \zeta = \mb F \mb L \mb \zeta = \mb L \mb F \mb L \mb L \mb \zeta = \cdots$.

Standard HMC can be viewed in terms of transitions on this ladder.  Additionally, we will see that this discrete state space view allows Equation \ref{eq fixed} to be solved directly by replacing the integral over all states with a short sum.  

\subsection{Standard HMC} \label{sec standard steps}
HMC as typically implemented consists of the following steps.  Here, $\mb \zeta^{\left( t, s \right)}$ represents the state at sampling step $t$, and sampling substep $s$.  Each numbered item below corresponds to a valid Markov transition for $p\left(\mb \zeta\right)$, satisfying detailed balance.  A full sampling step consists of the composition of all three Markov transitions.
\begin{enumerate}
\item \begin{enumerate}
  \item Generate a proposed update,
\begin{align}
\label{eq standard forward}
\mb \zeta' &= \mb F \mb L \zeta^{\left( t, 0 \right)}
.
\end{align}
On the state ladder in Figure \ref{fig op}b, this corresponds to moving up one rung ($\mb L$), and then moving from the right to the left side ($\mb F$).

  \item Accept or reject the proposed update using Metropolis-Hastings rules,
\begin{align}
\label{eq MH}
\pi_{accept} &= \min\left( 1, \frac{p\left(\mb \zeta' \right)}{p\left(\mb \zeta\right)} \right), \\
\zeta^{\left( t, 1 \right)} &= 
	\left\{\begin{array}{ccc}
		\mb \zeta' & & \text{with probability } \pi_{accept} \\
		\mb \zeta^{\left( t, 0 \right)} & & \text{with probability } 1 - \pi_{accept}
	\end{array}\right.
.
\end{align}
Note that since the transition $\mb F \mb L$ is its own inverse, the forward and reverse proposal distribution probabilities cancel in the Metropolis-Hastings rule in Equation \ref{eq MH}.

On rejection, the computations performed in Equation \ref{eq standard forward} are discarded.  In our new technique, this will no longer be true.
\end{enumerate}
  \item Flip the momentum,
	\begin{align}
		\mb \zeta^{\left( t, 2 \right)} &= \mb F \zeta^{\left( t, 1 \right)}.
	\end{align}
If the proposed update from Step 1 was accepted, then this moves $\zeta^{\left( t, 1 \right)}$ from the left back to the right side of the state ladder in Figure \ref{fig op}b, and prevents the trajectory from doubling back on itself.  If the update was rejected however, and $\zeta^{\left( t, 1 \right)}$ is already on the right side of the ladder, then this causes it to move to the left side of the ladder, and the trajectory to double back on itself.

Doubling back on an already computed trajectory is wasteful in HMC, both because it involves recomputing nearly redundant trajectories, and because the distance traveled before the sampler doubles back is the characteristic length scale beyond which HMC explores the state space by a random walk.
  \item Corrupt the momentum with noise,
	\begin{align}
		\mb \zeta^{\left( t+1, 0 \right)} &= \mb R\left(\beta\right) \zeta^{\left( t, 2 \right)}.
		\label{eq noise introduction}
	\end{align}
It is common to set $\beta = 1$, in which case the momentum is fully randomized every sampling step.  In our experiments (Section \ref{sec results}) however, we found that smaller values of $\beta$ produced large improvements in mixing time.  This is therefore a hyperparameter {that is probably worth} adjusting\footnote{One method for choosing $\beta$ \cite{Culpepper2011} which we have found to be effective is to set it such that it randomizes a fixed fraction $\alpha$ of the momentum per unit simulation time,
\begin{align}
	\beta &= \alpha^\frac{1}{\epsilon M}
	.
\end{align}
}.
\end{enumerate}

\section{Look Ahead HMC}
\label{LAHMC}

\begin{figure}
\begin{center}
\parbox[c]{0.95\linewidth}{
\begin{tabular}{l}
(a)	\includegraphics[width=0.75\linewidth]{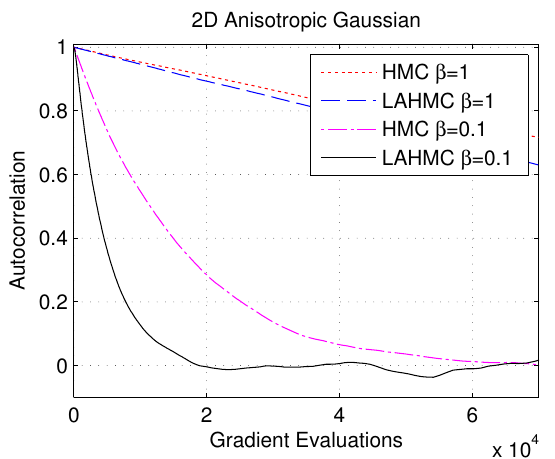}
\\
(b)	\includegraphics[width=0.75\linewidth]{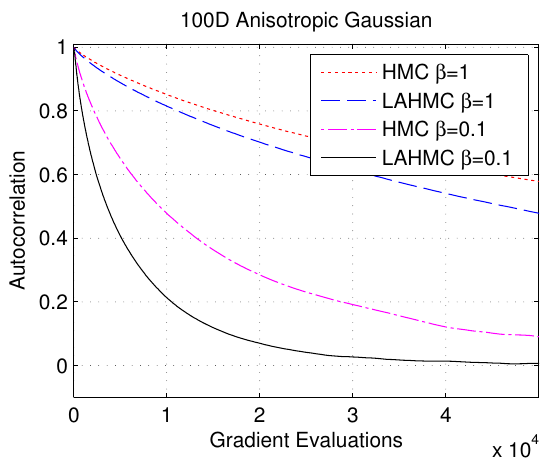}
\\
(c)	\includegraphics[width=0.75\linewidth]{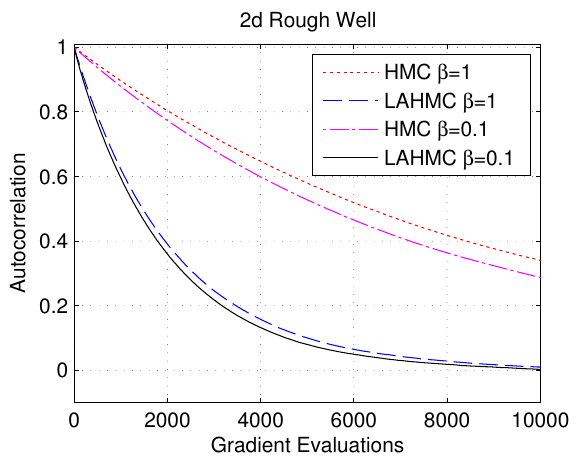}
\end{tabular}
} 
\end{center}
\caption{
Autocorrelation vs. number of function evaluations for standard HMC (no momentum randomization, $\beta = 1$), LAHMC with $\beta = 1$, persistent HMC ($\beta = 0.1$), and persistent LAHMC ($\beta = 0.1$) for {\em (a)} a two dimensional ill-conditioned Gaussian, {\em (b)} a one hundred dimensional ill-conditioned Gaussian, and {\em (c)} a two dimensional well conditioned energy function with a ``rough'' surface.  In all cases the LAHMC sampler demonstrates faster mixing.
}
\label{fig autocovariance}
\end{figure}

Here we introduce an HMC algorithm that relies on Markov transitions that do not obey detailed balance, but still satisfy the fixed point equation.  This algorithm eliminates much of the momentum flipping {that} occurs on rejection in HMC, and as a result greatly reduces random walk behavior.  It also prevents the trajectory computations {that} would typically be discarded on proposal rejection from being wasted.  We call our algorithm Look Ahead Hamiltonian Monte Carlo (LAHMC).

\subsection{Intuition}

In LAHMC, in situations {that} would correspond to a rejection in Step 1 of Section \ref{sec standard steps}, we will instead attempt to travel even farther by applying the leapfrog operator $\mb L$ additional times.  This section provides intuition for how this update rule was discovered, and how it can be seen to connect to standard HMC.  A more mathematically precise description will follow in the next several sections.

LAHMC can be understood in terms of a series of modifications of standard HMC.  The net effect of Steps 1 and 2 in Section \ref{sec standard steps} is to transition from state $\mb \zeta$ into either state $\mb L \mb \zeta$ or state $\mb F \mb \zeta$, depending on whether the update in Section \ref{sec standard steps} Step 1 was accepted or rejected.

We wish to minimize the transitions into state $\mb F \mb \zeta$.  In LAHMC we do this by replacing as many transitions from $\mb \zeta$ to $\mb F \mb \zeta$ as possible with transitions {that} instead go from $\mb \zeta$ to $\mb L^2 \mb \zeta$.  This would seem to change the number of transitions into both state $\mb F \mb \zeta$ and state $\mb L^2 \mb \zeta$, violating the fixed point equation.  However, the changes in incoming transitions from $\mb \zeta$ are exactly counteracted because the state $\mb F \mb L^2 \mb \zeta$ is similarly modified, so that it makes fewer transitions into the state $\mb L^2 \mb \zeta = \mb F \left(\mb F \mb L^2 \mb \zeta\right)$, and more transitions into the state $\mb F \mb \zeta = \mb L^2 \left(\mb F \mb L^2 \mb \zeta\right)$.

For some states, after this modification there will still be transitions between the states $\mb \zeta$ and $\mb F \mb \zeta$.  In order to further minimize these transitions, the process in the preceding paragraph is repeated for these remaining transitions and the state $\mb L^3 \mb \zeta$.  This process is then repeated again for states $\mb L^4 \mb \zeta$, $\mb L^5 \mb \zeta$, etc, up to some maximum number of leapfrog applications $K$.

\subsection{Algorithm} \label{sec alg}

LAHMC consists of the following two steps,
\begin{enumerate}
  \item Transition to a new state by applying the leapfrog operator $\mb L$ between 1 and $K \in \mc Z^+$ times, or by applying the momentum flip operator $\mb F$,
	\begin{align}
	\zeta^{\left( t, 1 \right)} &= 
		\left\{\begin{array}{ccc}
			\mb L \mb \zeta^{\left( t, 0 \right)} & & \text{with probability }  \pi_{\mb L^1}\left( \zeta^{\left( t, 0 \right)} \right) \\
			\mb L^2 \mb \zeta^{\left( t, 0 \right)} & & \text{with probability }  \pi_{\mb L^2}\left( \zeta^{\left( t, 0 \right)} \right) \\
			\cdots \\
			\mb L^K \mb \zeta^{\left( t, 0 \right)} & & \text{with probability }  \pi_{\mb L^K}\left( \zeta^{\left( t, 0 \right)} \right) \\
			\mb F \mb \zeta^{\left( t, 0 \right)} & & \text{with probability }  \pi_{\mb F}\left( \zeta^{\left( t, 0 \right)} \right) \\
		\end{array}\right.
	. \label{eq all transitions}
	\end{align}
	Note that there is no longer a Metropolis-Hastings accept/reject step.  The state update in Equation \ref{eq all transitions} is a valid Markov transition for $p\left( \mb \zeta \right)$ on its own.
  \item Corrupt the momentum with noise in an identical fashion as in Equation \ref{eq noise introduction},
	\begin{align}
		\mb \zeta^{\left( t+1, 0 \right)} &= \mb R\left(\beta\right) \zeta^{\left( t, 1 \right)}.
	\end{align}
\end{enumerate}

\subsection{Transition Probabilities}\label{sec transition}

We choose the probabilities $\pi_{\mb L^a}\left(\mb \zeta\right)$ for the leapfrog transitions from state $\mb \zeta$ to state $\mb L^a \mb \zeta$ to be
\begin{align}
\label{eq L^a prob}
	\pi_{\mb L^a}\left(\mb \zeta\right) &= \min\biggl[
		1 - \sum_{b < a} \pi_{\mb L^b}\left(\mb \zeta\right), \\ & \qquad \qquad \nonumber
		\frac{p\left( \mb F \mb L^a \mb \zeta \right)}{p\left( \mb \zeta \right)} \left( 1 - \sum_{b < a} \pi_{\mb L^b}\left(\mb F \mb L^a\mb \zeta\right) \right)
	\biggr]
.
\end{align}
Equation \ref{eq L^a prob} greedily sets the transition probability $\pi_{\mb L^a}\left(\mb \zeta\right)$ as large as possible, subject to the restrictions that the total transition probability out of state $\mb \zeta$ not exceed 1, and that the transition rate in the forward direction ($\mb \zeta \rightarrow \mb L^a \mb \zeta$) not exceed the transition rate in the reverse direction ($\mb F \mb L^a \mb \zeta \rightarrow \mb F \mb \zeta$)\footnote{
Although these transition probabilities do not satisfy detailed balance, as observed in \cite{Campos2014} they do satisfy an alternate condition sometimes used in physics which is known as {\em generalized} detailed balance. Generalized detailed balance does not lead to the same poor mixing behavior as detailed balance. Generalized detailed balance follows directly from the observation that
\begin{align}
\label{eq L^a prob relationship}
	p\left( \mb \zeta \right) \pi_{\mb L^a}\left(\mb \zeta\right) &= 
		p\left( \mb F \mb L^a \mb \zeta \right)
		\pi_{\mb L^a}\left(\mb F \mb L^a \mb \zeta\right)
.
\end{align}
}.

Any remaining unassigned probability is assigned to the momentum flip transition,
\begin{align}
	\pi_{\mb F}\left(\mb \zeta\right) &= 1 - \sum_{a}\pi_{\mb L^a}\left(\mb \zeta\right)
.
\end{align}

Note that transitions will be performed in a greedy fashion.  It is only necessary to compute the state $\mb L^a \mb \zeta$ and the transition probability $\pi_{\mb L^a}\left(\mb \zeta\right)$ if none of the transitions to states $\mb L^b \mb \zeta$, for $b < a$, have been taken.

\begin{table*}[t]
\begin{center}
\begin{tabular}{llccccc}
\large Distribution & \large Sampler & \large $\mb F \zeta$ & \large $\mb L \zeta$ & \large $\mb L^2 \zeta$ & \large $\mb L^3 \zeta$ & \large $\mb L^4 \zeta$ \\
\hline
\vspace{-1mm}\\
2d Gaussian & HMC $\beta = 1$ & 0.079 & 0.921 & 0 & 0 & 0  \\
2d Gaussian & LAHMC $\beta = 1$ & 0.000 & 0.921 & 0.035 & 0.044 & 0.000 \\
2d Gaussian & HMC $\beta = 0.1$ & 0.080 & 0.920 & 0 & 0 &  0 \\
2d Gaussian & LAHMC $\beta = 0.1$ & 0.000 & 0.921 & 0.035 & 0.044 & 0.000 \vspace{2mm}\\
100d Gaussian & HMC $\beta = 1$ & 0.147 & 0.853 & 0 & 0 & 0  \\
100d Gaussian & LAHMC $\beta = 1$ & 0.047 & 0.852 & 0.059 & 0.035 & 0.006 \\
100d Gaussian & HMC $\beta = 0.1$ & 0.147 & 0.853 & 0 & 0 & 0  \\
100d Gaussian & LAHMC $\beta = 0.1$ & 0.047 & 0.852 & 0.059 & 0.035 & 0.006 \vspace{2mm}\\
2d Rough Well & HMC $\beta = 1$ & 0.446 & 0.554 & 0 & 0 & 0  \\
2d Rough Well & LAHMC $\beta = 1$ & 0.292 & 0.554 & 0.099 & 0.036 & 0.019 \\
2d Rough Well & HMC $\beta = 0.1$ & 0.446 & 0.554 & 0 & 0 &  0 \\
2d Rough Well & LAHMC $\beta = 0.1$ & 0.292 & 0.554 & 0.100 & 0.036 & 0.019 \\
\end{tabular}
\end{center}
\caption{A table showing the fraction of transitions which occurred to each target state for the conditions plotted in Figure \ref{fig autocovariance}.  Note that LAHMC has far fewer momentum flips than standard HMC.
\label{tb endpoint}
}
\end{table*}
\begin{figure*}
\begin{center}
\parbox[c]{0.95\linewidth}{
\begin{tabular}{cc}
\begin{tabular}{cccc}
\includegraphics[width=0.22\linewidth]{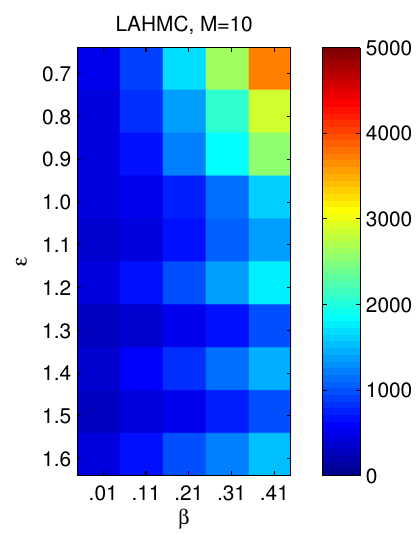} &
\includegraphics[width=0.22\linewidth]{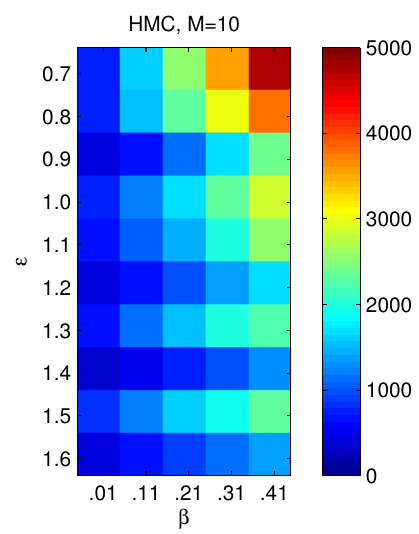} &
\includegraphics[width=0.22\linewidth]{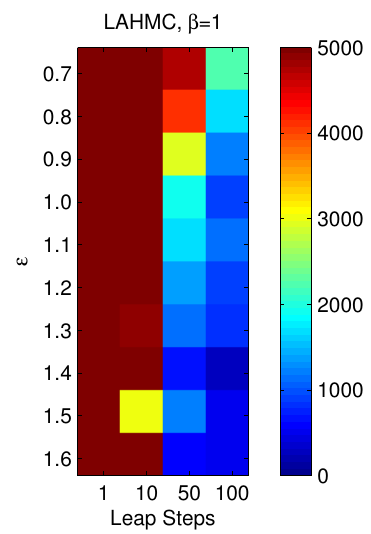} &
\includegraphics[width=0.22\linewidth]{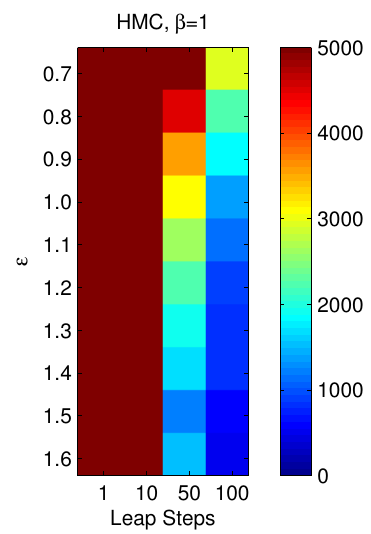} 
 \\
(a) & (b) & (c) & (d)
\end{tabular}
&
\end{tabular}
} 
\end{center}
\caption{
Images illustrating mixing time as a function of HMC hyperparameters for a two dimensional ill-conditioned Gaussian distribution.  Pixel intensity indicates the number of gradient evaluations required to reach an autocorrelation of 0.5.  LAHMC always outperforms HMC for the same hyperparameter settings.  
{\em (a)} LAHMC as a function of $\epsilon$ and $\beta$, for fixed $M = 10$, 
{\em (b)} HMC as a function of $\epsilon$ and $\beta$, for fixed $M = 10$, 
{\em (c)} LAHMC as a function of $\epsilon$ and $M$, for fixed $\beta = 1$, 
{\em (d)} HMC as a function of $\epsilon$ and $M$, for fixed $\beta = 1$.
}
\label{fig grid search}
\end{figure*}

\subsection{Fixed Point Equation}

We can substitute the transition rates from Section \ref{sec transition} into the left side of Equation \ref{eq fixed}, and verify that they satisfy the fixed point equation.  Note that the integral over all states is transformed into a sum over all source states from which transitions into state $\mb \zeta$ might be initiated. 
\begin{align}
	\int d\mb \zeta' & p\left( \mb \zeta' \right) T\left( \mb \zeta | \mb \zeta' \right) \nonumber \\
&=
	\int d\mb \zeta' p\left( \mb \zeta' \right) \biggl( \sum_a \pi_{\mb L^a}\left(\mb \zeta' \right) \delta\left( \mb \zeta - \mb L^a \mb \zeta' \right) \\ & \qquad \qquad \qquad \qquad \nonumber
	+ \pi_{\mb F}\left(\mb \zeta' \right) \delta\left( \mb \zeta - \mb F \mb \zeta' \right) \biggr)  , \\
&=
	\sum_a p\left( \mb L^{-a} \mb \zeta \right) \pi_{\mb L^a}\left(\mb L^{-a} \mb \zeta \right) 
	+ p\left( \mb F^{-1} \mb \zeta \right) \pi_{\mb F}\left(\mb F^{-1} \mb \zeta \right), \\
&=
	\sum_a p\left( \mb F \mb L^a \mb F \mb \zeta \right) \pi_{\mb L^a}\left(\mb F \mb L^a \mb F \mb \zeta \right) 
	+ p\left( \mb F \mb \zeta \right) \pi_{\mb F}\left(\mb F \mb \zeta \right), \\
&=
	\sum_a p\left( \mb F \mb \zeta \right) \pi_{\mb L^a}\left( \mb F \mb \zeta \right) 
	+ p\left( \mb F \mb \zeta \right) \pi_{\mb F}\left(\mb F \mb \zeta \right), \\
&=
	p\left( \mb F \mb \zeta \right) \left[
		\sum_a \pi_{\mb L^a}\left( \mb F \mb \zeta \right)
		+ \pi_{\mb F}\left(\mb F \mb \zeta \right)
	\right], \\
&=
	p\left( \mb \zeta \right)
.
\end{align}

\section{Experimental Results}
\label{sec results}

As illustrated in Figure \ref{fig autocovariance}, we compare the mixing time for our technique and standard HMC on three distributions.  HMC and LAHMC both had step length and number of leapfrog steps set to $\epsilon = 1$, and $M = 10$.  Values of $\beta$ were set to $1$ or $0.1$ as stated in the legend.  For LAHMC the maximum number of leapfrog applications was set to $K = 4$.  In all cases, LAHMC outperformed standard HMC for the same setting of hyperparameters, often by more than a factor of 2.  

The first two target distributions are 2 and 100 dimensional ill-conditioned Gaussian distributions.  In both Gaussians, the eigenvalues of the covariance matrix are log-linearly distributed between 1 and ${10}^6$.

The final target distribution was chosen to demonstrate that LAHMC is useful even for well conditioned distributions.  The energy function used was the sum of an isotropic quadratic and sinusoids in each of two dimensions,
\begin{align}
E\left( \mb x \right) &= \frac{1}{2\sigma_1^2} \left( x_1^2 + x_2^2\right) + \cos\left( \frac{\pi x_1}{\sigma_2} \right) + \cos\left( \frac{\pi x_2}{\sigma_2} \right)
,
\end{align}
where $\sigma_1 = 100$ and $\sigma_2 = 2$.  Although this distribution is well conditioned the sinusoids cause it to have a ``rough'' surface, such that traversing the quadratic well while maintaining a reasonable discretization error requires many leapfrog steps.

%

The fraction of the sampling steps resulting in each possible update for the samplers and energy functions in Figure \ref{fig autocovariance} is illustrated in Table \ref{tb endpoint}.  The majority of momentum flips in standard HMC were eliminated by LAHMC.  Note that the acceptance rate for HMC with these hyperparameter values is reasonably close to its optimal value of 65\% \cite{Neal:HMC}.

Figure \ref{fig grid search} shows several grid searches over hyperparameters for a two dimensional ill-conditioned Gaussian, and demonstrates that our technique outperforms standard HMC for all explored hyperparameter settings.  Due to computational constraints, the eigenvalues of the covariance of the Gaussian are 1 and $10^5$ in Figure \ref{fig grid search}, rather than $1$ and $10^6$ as in Figure \ref{fig autocovariance}a.


MATLAB and Python implementations of LAHMC are available at \url{http://github.com/Sohl-Dickstein/LAHMC}.  Figure \ref{fig autocovariance} and Table \ref{tb endpoint} can be reproduced by running generate\_figure\_2.m or generate\_figure\_2.py.

\section{Future Directions}

There are many powerful variations on standard HMC that are complementary to and could be combined naturally with the present work.  These include Riemann manifold HMC \cite{Girolami2011a}, quasi-Newton HMC \cite{zhang2011quasi}, Hilbert space HMC \cite{Beskos2011}, shadow Hamiltonian methods \cite{Izaguirre2004}, parameter adaptation techniques \cite{Wang2013}, Hamiltonian annealed importance sampling \cite{HAIS}, split HMC \cite{Shahbaba2011}, and tempered trajectories \cite{Neal:HMC}.

It should be possible to further reduce random walk behavior by exploring new topologies and allowed state transitions.  Two other schemes have already been explored, though with only marginal benefit.  In one scheme as many flips as possible are replaced by identity transitions.  This is described in the note \cite{Sohl-Dickstein2012}.  In a second scheme, a state space is constructed with two sets of auxiliary momentum variables, and an additional momentum-swap operator which switches the two momenta with each other is included in the allowed transitions.  In this scenario, in situations {that} would typically lead to momentum flipping, with high probability the two sets of momenta can instead be exchanged with each other.  This leads to momentum randomization on rejection, rather than momentum reversal.  Unfortunately, though this slightly improves mixing time, it still amounts to a random walk on a similar length scale.  The exploration of other topologies and allowed transitions will likely prove fruitful.

Any deterministic, reversible, discrete stepped trajectory through a state space can be mapped onto the ladder structure in Figure \ref{fig op}. The Markov transition rules presented in this paper could therefore be applied to a wide range of problems.  All that is required in addition to the mapping is an auxiliary variable indicating direction along that trajectory.  In HMC, the momentum variable doubles as a direction indicator, but there could just as easily be an additional variable $d \in \{-1,1\}$, $p\left(d= 1\right) = \frac{1}{2}$, which indicates whether transitions are occurring up or down the ladder.  The efficiency of the exploration then depends only on choosing a sensible, approximately energy conserving, trajectory.


\section*{Acknowledgments} 

We would like to thank Bruno Olshausen and the members of the Redwood Center for Theoretical Neuroscience at Berkeley and Ganguli lab at Stanford for many thoughtful discussions and for encouragement during the course of this work. We would also like to thank the anonymous reviewers for a careful reading of the text, and thoughtful and actionable feedback. MRD and JSD were both supported by NSF grant IIS-1219199 to Michael R. DeWeese. JSD was additionally supported by the Khan Academy. MM was supported by NGA grant HM1582-081-0007 to Bruno Olshausen and NSF grant IIS-1111765 to Bruno Olshausen. This material is based upon work supported in part by the U.S. Army Research Laboratory and the U.S. Army Research Office under contract number W911NF-13-1-0390.

%


\bibliography{icml2014}
\bibliographystyle{icml2014}

\end{document}